\documentclass[groupedaddress, superscriptaddress, citesort]{revtex4}
\usepackage{amssymb, amsmath}
\usepackage[dvips]{graphicx}

\def\be{\begin{equation}}
\def\ee{\end{equation}}
\def\ba{\begin{array}}
\def\ea{\end{array}}

\begin{document}
\baselineskip=13pt

\title {Super-quantum correlation and geometry for Bell-diagonal states with weak measurements}
\author{Yao-Kun Wang}
\affiliation{School of Mathematical Sciences,  Capital Normal
University,  Beijing 100048,  China}
\affiliation{College of Mathematics,  Tonghua Normal University, Tonghua 134001,  China}
\author{Teng Ma}
\affiliation{School of Mathematical Sciences,  Capital Normal University,  Beijing 100048,  China}
\author{Heng Fan}
\affiliation{Institute of Physics, Chinese Academy of Sciences, Beijing 100190, China}
\author{Shao-Ming Fei}
\affiliation{School of Mathematical Sciences, Capital Normal University, Beijing 100048, China}
\affiliation{Max-Planck-Institute for Mathematics in the Sciences, 04103 Leipzig, Germany}
\author{Zhi-Xi Wang}
\affiliation{School of Mathematical Sciences,  Capital Normal
University,  Beijing 100048,  China}

\begin{abstract}
We propose ``weak one-way deficit'' by weak measurements as the generalization of one-way deficit defined for standard
projective measurements. The weak one-way deficit for Werner state is obtained analytically.
We find that weak one-way deficit is smaller than the standard one-way deficit, which contrasts with a
straightforward expectation based on the known fact that super-quantum discord by weak measurement
is always larger than the quantum discord defined by projective measurement.
On the other hand,
by tuning the weak measurement continuously to the projective measurement, both weak one-way deficit and super-quantum
discord converge to the same value, which is either the one-way deficit or the quantum discord both quantifying
quantum correlation. In this sense, weak measurement does not necessarily capture more quantumness of correlations.
We also give the geometry of super-quantum discord of the Bell-diagonal states with explicit geometrical figures.
As an application, the dynamic behavior of super-quantum correlation including super-quantum discord and weak one-way
deficit under decoherence is investigated.
We find that the order relation of the super-quantum correlation and the quantum correlation keep unchanged under the phase flipping channel for the Bell-diagonal states and the Werner states.
\end{abstract}

\maketitle

\section{Introduction}
Weak measurement introduced by Aharonov, Albert, and Vaidman (AAV) \cite{Aharonov} in 1988 is universal in the sense that any generalized measurements can be realized as a sequence of weak measurements
which result in small changes to the quantum state for all outcomes \cite{Oreshkov}.
Weak measurement is very useful, much progress has been made in this interesting field, including the contribution  of the probe dynamics to the weak measurement \cite{Lorenzo}, such as weak
measurement with arbitrary probe \cite{Johansen}, entangled probes \cite{Menzies}, a qubit probe \cite{Wu},
weak measurement with a spin observable \cite{duck,Pan}, and so on.
In addition, weak measurement realized by some experiments is also very useful for high-precision measurements.
For example, Hosten and Kwiat \cite{Hosten} use the weak measurement to observe the spin Hall effect in light;
Dixon \emph{et al.} \cite{Dixon} apply the weak measurement to detect very small transverse beam  deflections;
Gillett \emph{et al.} \cite{Gillett} study the weak measurement to examine the feedback control of quantum systems in the presence of noise.

 The quantum correlations of quantum states include entanglement and other kinds of nonclassical correlations.
  It is well accepted that the quantum correlations are more general than entanglement \cite{bennett, zurek1,Henderson}.
  A canonical measure of quantum correlations is the quantum discord \cite{Ollivier} which describes
   the quantumness of correlations. It quantifies how much a system can be disrupted when we observe it to obtain the classical information.
Remarkable developments have been achieved toward
the importance and applications of quantum discord. In particular, there are some precise expressions for quantum discord for two-qubit states, such as
 for the $X$ states \cite{Ali,Li,chen,shi,Vinjanampathy}. The geometry of the quantum correlations about those states is also investigated \cite{Li,langcaves}.
 Recently, researches on the dynamics of quantum discord in various noisy environments have revealed many attractive features \cite{werlang}.
 It is demonstrated that discord is more robust than entanglement for both Markovian and non-Markovian dissipative processes.

Besides quantum discord, a lot of other measures of quantum correlations have been given, such as the quantum deficit \cite{oppenheim,horodecki}, measurement-induced disturbance \cite{luo}, symmetric discord \cite{piani,wu}, relative entropy of discord and dissonance \cite{modi}, geometric discord \cite{luoandfu,dakic}, and continuous-variable discord \cite{adesso,giorda}. A nice review paper about quantum correlations can be found in \cite{modi2}. Among them, the work deficit \cite{oppenheim} is one operational approach to quantify quantum correlations. From the physical point of view, quantum deficit originates in describing a process which tries to extract work by nonlocal operation from a correlated system coupled to a heat bath in the case of pure states.
It can be related to more general forms of quantum correlations. Similarly, quantum discord
can also be justified by a physical interpretation \cite{zurek2} by that the erasure of quantum correlations must lead to entropy production in the system and the environment \cite{Merali}. Similar to quantum discord, Oppenheim \emph{et al.} propose a definition of work deficit and verify that quantum deficit is equal to the difference of the mutual information and classical deficit \cite{oppenheim2}. Some simple forms of deficit are found in \cite{horodecki2}.
It is shown that the classical information deficit \cite{synak1} is lower-bounded by the (regularized)
relative entropy of entanglement for Werner and isotropic states \cite{synak2}.
Recently, the definition of the one-way deficit is given
by the relative entropy over all local von Neumann measurements on one subsystem which reveals the fundamental role of quantum correlations as a resource for the distribution of entanglement \cite{Streltsov0,chuan}. Other definitions of the one-way deficit by von Neumann measurements on one subsystem are also
proposed \cite{streltsov}.

Quantum discord and the one-way deficit both are quantum correlations based on von Neumann measurement.
Since the fundamental role of weak measurements, it is interesting to know
how those quantum correlations will be for weak measurement?  Recently, it is shown that weak measurement performed on one of the subsystems can lead to ``super-quantum discord'' that is always larger than the normal quantum discord captured by the strong (projective) measurements \cite{singh}.
It is natural to ask whether weak measurements can always capture more quantumness of correlations.
In this article, we propose a definition of the weak one-way deficit by weak measurement.
Interestingly, by tuning continuously from
strong measurement to weak measurement, the discrepancy between the super-quantum discord and weak one-way deficit
becomes larger. In comparison with super-quantum discord which is larger than the standard discord,
the weak one-way deficit decreases for weak measurement, while they are completely the same for projective measurement. In this sense, weak measurement does
not always capture more quantumness of correlations. It depends on the specified measure of quantum correlations.
We calculate the weak one-way deficit for Werner states and compare super-quantum correlation with quantum correlation in Sect. \ref{II}.
We give super-quantum discord for Bell-diagonal states and depict the level surface of constant super-quantum discord and quantun discord in Sect. \ref{III}.
In Sect. \ref{IIII}, the dynamic behavior of super-quantum discord and weak one-way deficit under decoherence is investigated.
A brief conclusion is given in Sect. \ref{IIIII}.

\section{Super-quantum correlation of Werner State with Weak Measurements}\label{II}

The quantum discord for bipartite quantum state \(\rho_{AB}\) with the strong measurement \(\{\Pi^B_i\}\) performed on the subsystem $B$ is the difference between the mutual information $I(\rho_{AB})$ \cite{partovi} and classical correlation $J_B(\rho_{AB})$ \cite{Henderson}
\begin{equation}
\label{dis}
D(\rho_{AB})=\min_{\{\Pi_i^B\}} \sum_i p_i S(\rho_{A|i})+S(\rho_{B})-S(\rho_{AB})
\end{equation}
with the minimization going over all projection-valued measurements \(\{\Pi^B_i\}\),
where \(S(\rho) = - \mbox{tr} \left(\rho \log \rho\right)\) is the von Neumann entropy of a quantum state \(\rho\), \(\rho_B\) is the reduced density matrices of \(\rho_{AB}\) and
\begin{equation}
p_i = \mbox{tr}_{AB}[(I_A \otimes \Pi^B_i ) \rho_{AB} ( {I}_A \otimes \Pi^B_i) ],\
\rho_{A|i} = \frac{1}{p_i} \mbox{tr}_B[({I}_A \otimes \Pi^B_i) \rho_{AB} ({I}_A \otimes \Pi^B_i)].
\end{equation}

The weak measurement operators are given by \cite{Oreshkov}
\begin{eqnarray}
 P(x) &=& \sqrt{\frac{(1-\tanh x)}{2}} \Pi_0 + \sqrt{ \frac{(1+\tanh x)}{2}} \Pi_1,  \nonumber\\
 P(-x) & = & \sqrt{\frac{(1+\tanh x)}{2}}\Pi_0 + \sqrt{\frac{(1-\tanh x)}{2}}\Pi_1, \nonumber\\
\end{eqnarray}
where $x$ is the measurement strength parameter, $\Pi_0$ and $\Pi_1$ are two orthogonal projectors with $\Pi_0 + \Pi_1 =I$. The weak measurement operators satisfy: (i) $P^{\dagger}(x)P(x) + P^{\dagger}(-x)P(-x) = I$, (ii) $\lim_{x \rightarrow \infty} P(x) = \Pi_0$ and  $\lim_{x \rightarrow \infty} P(-x) = \Pi_1$.

Recently, Singh and Pati propose the super-quantum discord for bipartite quantum state \(\rho_{AB}\) with weak measurement on the subsystem $B$ \cite{singh},
the super-quantum discord denoted by $D_w(\rho_{AB})$ is given by
\begin{equation}
D_w(\rho_{AB})=  \min_{\{\Pi_i^B\}}  S_w(A|\{P^{B}(x)\})+S(\rho_{B})-S(\rho_{AB})\label{sqdiscord}
\end{equation}
with the minimization going over all projection-valued measurements \(\{\Pi^B_i\}\),
where \(S(\rho) = - \mbox{tr} \left(\rho \log \rho\right)\) is the von Neumann entropy of a quantum state \(\rho\),
 \(\rho_B\) is the reduced density matrices of \(\rho_{AB}\), and
\begin{equation}
 S_w(A|\{P^{B}(x)\})= p(x) S(\rho_{A|P^{B}(x)}) + p(-x) S(\rho_{A|P^{B}(-x)}),
\end{equation}
\begin{equation}
 p(\pm x) =\mbox{tr}_{AB}[(I \otimes P^{B}(\pm x)) \rho_{AB} (I \otimes P^{B}(\pm x))],\label{probab}
\end{equation}
\begin{equation}
 \rho_{A|P^{B}(\pm x)}=\frac{\mbox{tr}_{B}[(I \otimes P^{B}(\pm x)) \rho_{AB} (I \otimes  P^{B}(\pm x))]}
{\mbox{tr}_{AB}[(I \otimes P^{B}(\pm x)) \rho_{AB} (I \otimes P^{B}(\pm x))]}, \label{state1}
\end{equation}
$\{P^{B}(x)\}$ is weak measurement operators performed on the subsystem $B$.

Streltsov \emph{et al.} give the definition of the one-way deficit by von Neumann measurements on the subsystem $B$  \cite{streltsov}
\begin{eqnarray}
\Delta^{\rightarrow}(\rho_{AB})=\min\limits_{\{\Pi_{k}\}}S(\sum\limits_{i}\Pi_{k}\rho_{AB}\Pi_{k})-S(\rho_{AB}).\label{definition}
\end{eqnarray}

Now, let us define what we call as the weak one-way deficit by weak measurement on the subsystem $B$,
\begin{eqnarray}
\Delta^{\rightarrow}_{w}(\rho_{AB})=\min\limits_{\{\Pi_{k}\}}S(\sum\limits_{P(\pm x)}(I \otimes P^{B}(\pm x)) \rho_{AB} (I \otimes P^{B}(\pm x)))-S(\rho_{AB}).\label{owdeficit}
\end{eqnarray}

The Werner state is a special case of Bell-diagonal state and
it can be a maximally entangled state in a special case, namely
\begin{equation}
\rho_{AB} = z |\Psi^-\rangle\langle\Psi^-|+\frac{(1-z)}{4}I,
\end{equation}
where $|\Psi^-\rangle=(|01\rangle - |10\rangle)/\sqrt{2}$.

For the Bell-diagonal states, quantum discord equals to the one-way deficit \cite{Wang},
which is also true for the Werner state. Quantum discord for Werner state is given by \cite{luo}
\begin{eqnarray}
D(\rho_{AB})=\frac{1-z}{4}\log(1-z)-\frac{1+z}{2}\log(1+z)+\frac{1+3z}{4}\log(1+3z).
\end{eqnarray}
And the super-quantum discord is given by \cite{singh}
\begin{eqnarray}
D_w(\rho_{AB}) &=&\frac{3(1-z)}{4}\log\left(\frac{1-z}{4}\right)+\frac{(1+3z)}{4}\log\left(\frac{1+3z}{4}\right)\nonumber\\
& &+1-[\frac{(1-z\tanh x)}{2}\log\left(\frac{1-z\tanh x}{2}\right)\nonumber\\
& &+\frac{(1+z\tanh x)}{2}\log\left(\frac{1+z\tanh x}{2}\right)].
\end{eqnarray}

Next, we will evaluate the weak one-way deficit defined above
for Werner state. Since Werner state is rotationally invariant, therefore, one gets the same result for the entropy of the state after measurement for any
measurement basis. Hence, one does not need to do minimization over all measurement bases. In fact, the $d\times d$-dimensional bipartite Werner state satisfies $\rho=(U\otimes U)\rho(U^{\dagger}\otimes U^{\dagger})$ for all unitary operators $U$ acting on $d$-dimensional Hilbert space, and such unitary transformations do not change the entropy of the state. Let $\Pi_0=|0\rangle\langle0|, \Pi_1=|1\rangle\langle1|$, the entropy of the post-measurement state is given by
\begin{eqnarray}
& &S(\sum\limits_{P(\pm x)}(I \otimes P^{B}(\pm x))\rho(I \otimes P^{B}(\pm x)))\nonumber\\
&=&S(\sum\limits_{P(\pm x)}(U \otimes P^{B}(\pm x)U)\rho(U^{\dagger}\otimes U^{\dagger}P^{B}(\pm x)))\nonumber\\
&=&S((U^{\dagger}\otimes U^{\dagger})\sum\limits_{P(\pm x)}(U \otimes P^{B}(\pm x)U)\rho(U^{\dagger}\otimes U^{\dagger}P^{B}(\pm x))(U\otimes U))\nonumber\\
&=&S(\sum\limits_{P(\pm x)}(I\otimes U^{\dagger}P^{B}(\pm x)U)\rho(I\otimes U^{\dagger}P^{B}(\pm x)U))\nonumber\\
&=&S(\sum\limits_{P(\pm x)}(I\otimes (\sqrt{\frac{(1\mp\tanh x)}{2}}U^{\dagger}\Pi_0U +\sqrt{\frac{(1\pm\tanh x)}{2}}U^{\dagger}\Pi_1U))\nonumber\\
& &\qquad\quad\times\rho(I\otimes (\sqrt{\frac{(1\mp\tanh x)}{2}}U^{\dagger}\Pi_0U + \sqrt{\frac{(1\pm\tanh x)}{2}}U^{\dagger}\Pi_1U)))\nonumber\\
\end{eqnarray}
and
\begin{eqnarray}
& &\min\limits_{\{\Pi_{k}\}}S(\sum\limits_{P(\pm x)}(I \otimes P^{B}(\pm x)) \rho (I \otimes P^{B}(\pm x)))\nonumber\\
&=&S(\sum\limits_{P(\pm x)}(I\otimes (\sqrt{\frac{(1\mp\tanh x)}{2}}|0\rangle\langle0|+\sqrt{\frac{(1\pm\tanh x)}{2}}|1\rangle\langle1|)\cdot\nonumber\\
& &\rho(I\otimes (\sqrt{\frac{(1\mp\tanh x)}{2}}|0\rangle\langle0| + \sqrt{\frac{(1\pm\tanh x)}{2}}|1\rangle\langle1|)))\nonumber\\
&=&S(\frac{-z}{2\cosh x}|01\rangle\langle10|+\frac{1+z}{4}|10\rangle\langle10|+\frac{1-z}{4}|00\rangle\langle00|\nonumber\\
& &+\frac{1+z}{4}|01\rangle\langle01|+\frac{-z}{2\cosh x}|10\rangle\langle01|+\frac{1-z}{4}|11\rangle\langle11|).
\end{eqnarray}

The eigenvalues of the post-measurement state are $\lambda_1=\frac{1+z}{4}+\frac{z}{2\cosh x}, \lambda_2=\frac{1+z}{4}-\frac{z}{2\cosh x},
\lambda_{3,4}=\frac{1-z}{4}$. By Eq.(\ref{owdeficit}), the weak one-way deficit of the Werner state is
\begin{eqnarray}
\Delta^{\rightarrow}_{w}(\rho_{AB})&=&\frac{1+3z}{4}\log\left(\frac{1+3z}{4}\right)+\frac{1-z}{4}\log\left(\frac{1-z}{4}\right)\nonumber\\
& &-\left(\frac{1+z}{4}+\frac{z}{2\cosh x}\right) \log\left(\frac{1+z}{4}+\frac{z}{2\cosh x}\right)\nonumber\\
& &-\left(\frac{1+z}{4}-\frac{z}{2\cosh x}\right) \log\left(\frac{1+z}{4}-\frac{z}{2\cosh x}\right).
\end{eqnarray}

In Fig.1, we plot the super-quantum correlation and quantum correlation of the Werner state. We find that super-quantum discord is larger than quantum discord (one-way deficit), and the one-way deficit is larger than weak one-way deficit(i.e., $\Delta^{\rightarrow}_{w}<\Delta^{\rightarrow}=D<D_w$). Weak one-way deficit approaches to zero for smaller values of $x$. super-quantum discord and weak one-way deficit approach to quantum discord (one-way deficit) for larger values of $x$.

 Therefore, for the case of strong measurements, i.e., $x \rightarrow \infty $, we have $\Delta^{\rightarrow}_{w}=\Delta^{\rightarrow}=D=D_w$. It has been proved that super-quantum discord is greater than the normal discord \cite{singh}. But weak one-way deficit is less than the one-way deficit. This is due to that the weak measurements disturb the subsystem of a composite system weakly, so that the entropy of composite system does not change much. It cannot capture more quantum correlation than the von Neumann measurement. As a result, weak measurements do not always reveal more quantumness.

\begin{figure}[h]
\raisebox{17em}{(a)}\includegraphics[width=6.25cm]{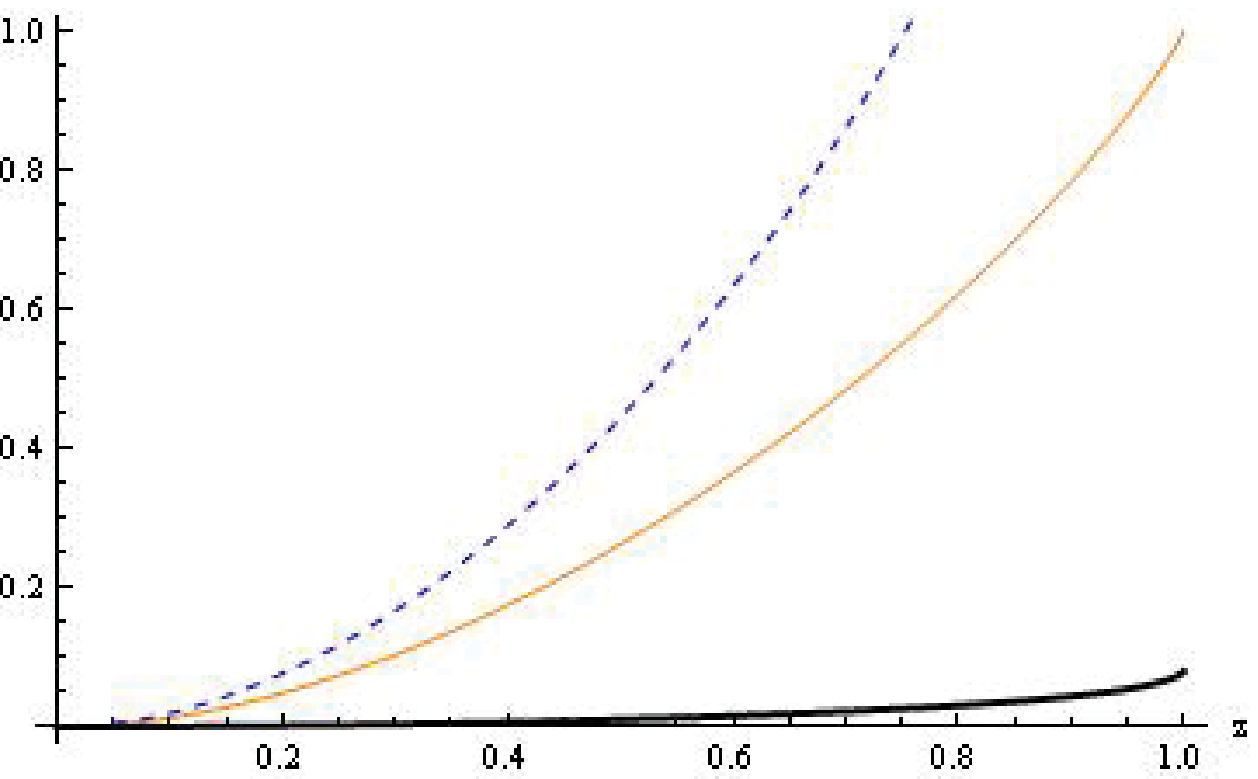}
\raisebox{17em}{(b)}\includegraphics[width=6.25cm]{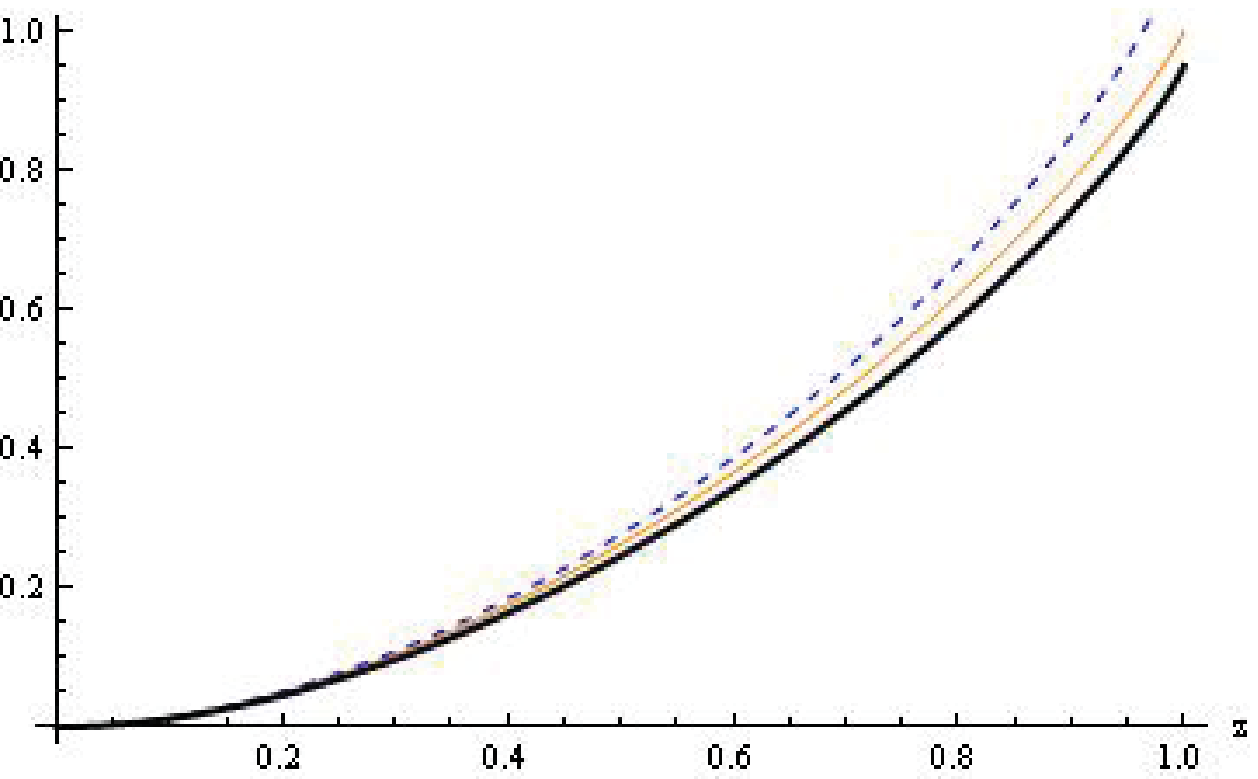}
\label{Fig:1}
\end{figure}
\begin{figure}[h]
\begin{center}
\caption{(Color online) Super-quantum correlation and quantum correlation for the Werner state as a function of $z$: super-quantum discord (dashed blue line), quantum discord (one-way deficit) (solid orange line) and weak one-way deficit (thick black line), (a) $x=0.2$, (b) $x=2$.}
\end{center}
\end{figure}

\section{Geometrical depiction of super-quantum discord for Bell-diagonal states with weak measurements }\label{III}
Now we evaluate the super-quantum discord for two-qubit Bell-diagonal state,
\begin{eqnarray}
\rho_{AB}=\frac{1}{4}(I\otimes I+\sum_{i=1}^3c_i\sigma_i\otimes\sigma_i).
\end{eqnarray}
The eigenvalues of $\rho_{AB}$ are
\begin{eqnarray}
\lambda_5=\frac{1}{4}(1-c_{1}-c_{2}-c_{3}),~~ \lambda_6=\frac{1}{4}(1-c_{1}+c_{2}+c_{3}),\nonumber\\
\lambda_7=\frac{1}{4}(1+c_{1}-c_{2}+c_{3}),~~ \lambda_8=\frac{1}{4}(1+c_{1}+c_{2}-c_{3}).\nonumber
\end{eqnarray}
The entropy of  $\rho_{AB}$ is
\begin{eqnarray}
S(\rho_{AB})
&=&-\sum_{i=5}^{8}\lambda_{i}\log \lambda_i\nonumber\\
&=&2-\frac{1-c_1-c_2-c_3}{4}\log (1-c_1-c_2-c_3)\nonumber\\
& &-\frac{1-c_1+c_2+c_3}{4}\log (1-c_1+c_2+c_3)\nonumber\\
& &-\frac{1+c_1-c_2+c_3}{4}\log (1+c_1-c_2+c_3)\nonumber\\
& &-\frac{1+c_1+c_2-c_3}{4}\log (1+c_1+c_2-c_3). \label{entropy}
\end{eqnarray}
Let $\{\Pi_{k}=|k\rangle\langle k|, k=0, 1\}$
be the local measurement for the part $B$ along the computational base ${|k\rangle}$. Then, any weak measurement operators for the part $B$ can be written as
\begin{eqnarray}
I\otimes P(\pm x)=\sqrt{\frac{(1\mp\tanh x)}{2}}I\otimes V\Pi_0V^{\dag} + \sqrt{\frac{(1\pm\tanh x)}{2}}I\otimes V\Pi_1 V^{\dag},
\end{eqnarray}
for some unitary $V\in U(2)$. But any unitary $V$ can be written, up to a constant phase, as $V=tI+i\vec{y}\cdot\vec{\sigma}$
with $t\in R$, $\vec{y}=(y_{1}, y_{2}, y_{3})\in R^{3}$, and $t^{2}+y_{1}^{2}+y_{2}^{2}+y_{3}^{2}=1. $

After the weak measurement, the state $\rho_{AB}$ will change to the ensemble $\{\rho_{A|P^{B}(\pm x)},\, p(\pm x)\}$.
We need to evaluate $\rho_{A|P^{B}(\pm x)}$ and $p(\pm x)$. By using the relations \cite{luo},
\begin{eqnarray}
V^{\dag}\sigma_{1}V=(t^{2}+y_{1}^{2}-y_{2}^{2}-y_{3}^{2})\sigma_{1}+2(ty_{3}+y_{1}y_{2})
\sigma_{2}+2(-ty_{2}+y_{1}y_{3})\sigma_{3},\nonumber\\[2mm]
V^{\dag}\sigma_{2}V=2(-ty_{3}+y_{1}y_{2})\sigma_{1}+(t^{2}+y_{2}^{2}-y_{1}^{2}-y_{3}^{2})
\sigma_{2}+2(ty_{1}+y_{2}y_{3})\sigma_{3},\nonumber\\[2mm]
V^{\dag}\sigma_{3}V=2(ty_{2}+y_{1}y_{3})\sigma_{1}+2(-ty_{1}+y_{2}y_{3})
\sigma_{2}+(t^{2}+y_{3}^{2}-y_{1}^{2}-y_{2}^{2})\sigma_{3},\nonumber
\end{eqnarray}
and
$\Pi_{0}\sigma_{3}\Pi_{0}=\Pi_{0}$, $\Pi_{1}\sigma_{3}\Pi_{1}=-\Pi_{1}$, $\Pi_{j}\sigma_{k}\Pi_{j}=0$
for $j=0,1$, $k=1, 2$, from Eqs. (\ref{probab}) and (\ref{state1}), we obtain $p(\pm x) =\frac{1}{2}$ and
\begin{eqnarray}
\rho_{A|P^{B}(+x)}=\frac{1}{2}(I-\tanh x(c_{1}z_{1}\sigma_{1}+c_{2}z_{2}\sigma_{2}+c_{3}z_{3}\sigma_{3})),\nonumber\\
\rho_{A|P^{B}(-x)}=\frac{1}{2}(I+\tanh x(c_{1}z_{1}\sigma_{1}+c_{2}z_{2}\sigma_{2}+c_{3}z_{3}\sigma_{3})),
\end{eqnarray}
where $z_{1}=2(-ty_{2}+y_{1}y_{3})$, $z_{2}=2(ty_{1}+y_{2}y_{3})$ and $z_{3}=t^{2}+y_{3}^{2}-y_{1}^{2}-y_{2}^{2}$.

Set $\theta=\sqrt{|c_{1}z_{1}|^{2}+|c_{2}z_{2}|^{2}+|c_{3}z_{3}|^{2}}$. Then,
\begin{eqnarray}
S(\rho_{A|P^{B}(+x)})=S(\rho_{A|P^{B}(-x)})=-\frac{1-\theta\tanh x}{2}\log\frac{1-\theta\tanh x}{2}-\frac{1+\theta\tanh x}{2}\log\frac{1+\theta\tanh x}{2}
\end{eqnarray}
and
\begin{eqnarray}
 S_w(A|\{P^{B}(x)\})&=&\frac{1}{2}S(\rho_{A|P^{B}(x)})+\frac{1}{2}S(\rho_{A|P^{B}(-x)})\nonumber\\
 &=& -\frac{1-\theta\tanh x}{2}\log\frac{1-\theta\tanh x}{2}-\frac{1+\theta\tanh x}{2}\log\frac{1+\theta\tanh x}{2}.
\end{eqnarray}

Let $c=\max\{|c_{1}|, |c_{2}|, |c_{3}|\},$ then $\theta\leq\sqrt{|c|^{2}(|z_{1}|^{2}+|z_{2}|^{2}+|z_{3}|^{2})}=c$. Hence we get
\begin{eqnarray}
\sup\limits_{\{V\}}\theta=c.
\end{eqnarray}
The range of values allowed for $\theta$ is $\theta\in[0, c]$.  It can be verified that $S_w(A|\{P^{B}(x)\})$
is a monotonically decreasing function of $\theta$ in the interval of $[0, c]$.  The minimal value of $S_w(A|\{P^{B}(x)\}$ can be attained at point $c$,
\begin{equation}
\min_{\{\Pi_i^B\}}  S_w(A|\{P^{B}(x)\})=-\frac{1-c\tanh x}{2}\log\frac{1-c\tanh x}{2}-\frac{1+c\tanh x}{2}\log\frac{1+c\tanh x}{2}.
\end{equation}
Then, by Eq. (\ref{sqdiscord}), (\ref{entropy}) and $S(\rho_{B})=1$, the
super-quantum discord of Bell-diagonal states is given by
\begin{eqnarray}
D_w(\rho _{AB})&=&\min_{\{\Pi_i^B\}}S_w(A|\{P^{B}(x)\}) - S(A|B)\nonumber\\
&=&\frac{1-c_1-c_2-c_3}{4}\log (1-c_1-c_2-c_3)+\frac{1-c_1+c_2+c_3}{4}\log (1-c_1+c_2+c_3)\nonumber\\
& &+\frac{1+c_1-c_2+c_3}{4}\log (1+c_1-c_2+c_3)+\frac{1+c_1+c_2-c_3}{4}\log (1+c_1+c_2-c_3)\nonumber\\
& &-\frac{1-c\tanh x}{2}\log(1-c\tanh x)-\frac{1+c\tanh x}{2}\log(1+c\tanh x).
\end{eqnarray}

\begin{figure}[h]
\raisebox{17em}{(a)}\includegraphics[width=6.25cm]{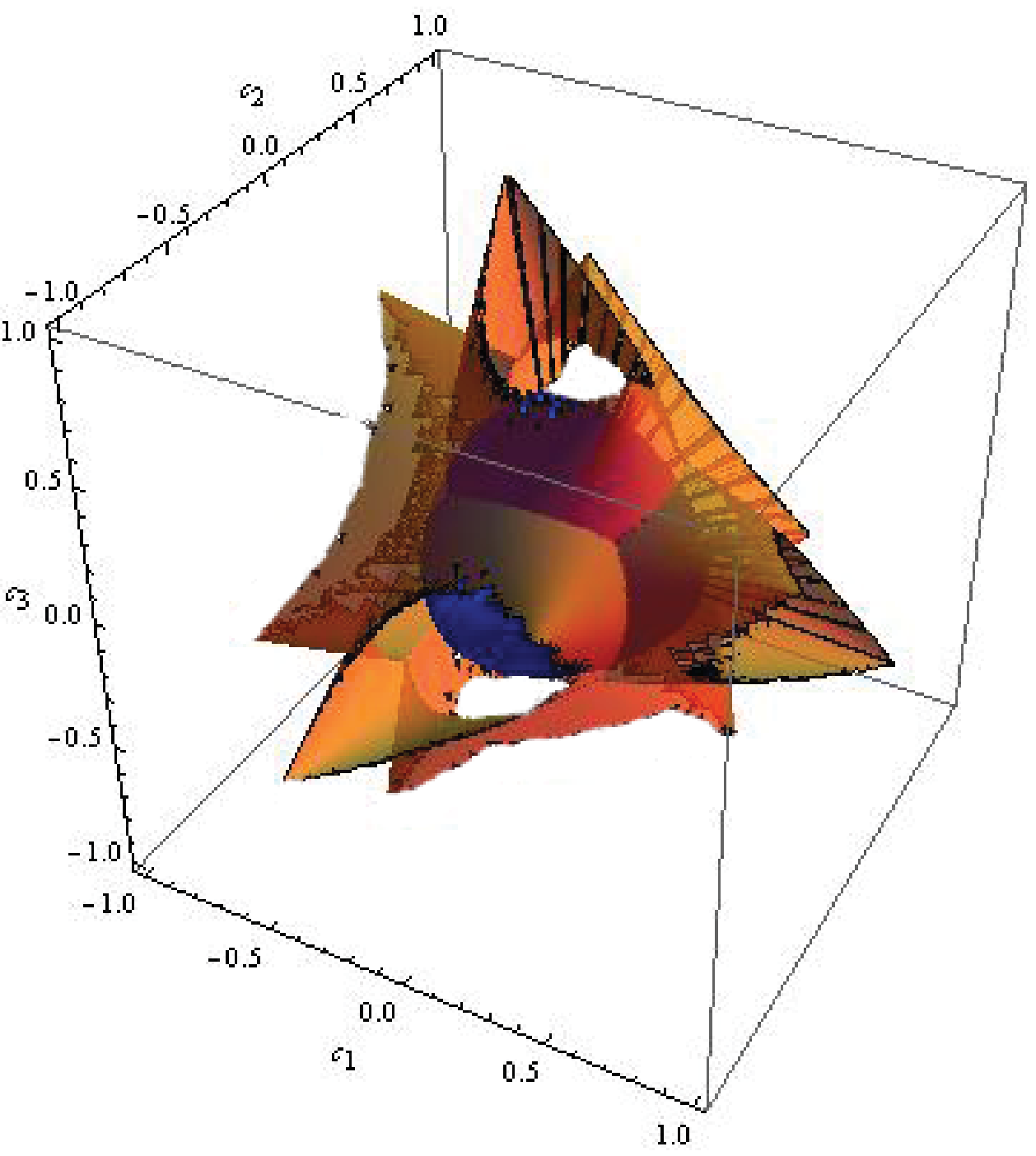}
\qquad
\raisebox{17em}{(b)}\includegraphics[width=6.25cm]{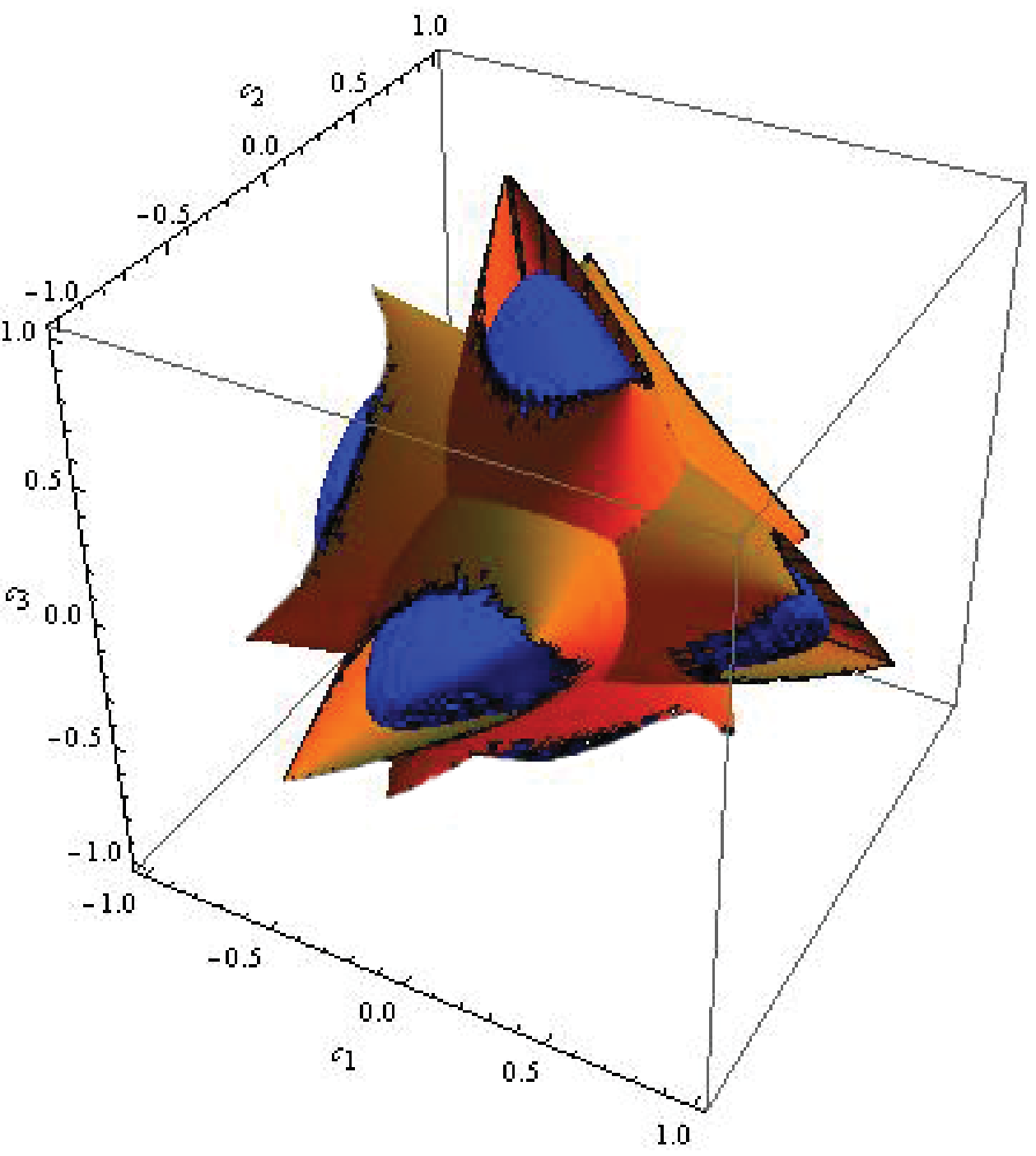}
\begin{center}
\raisebox{17em}{(c)}\includegraphics[width=6.25cm]{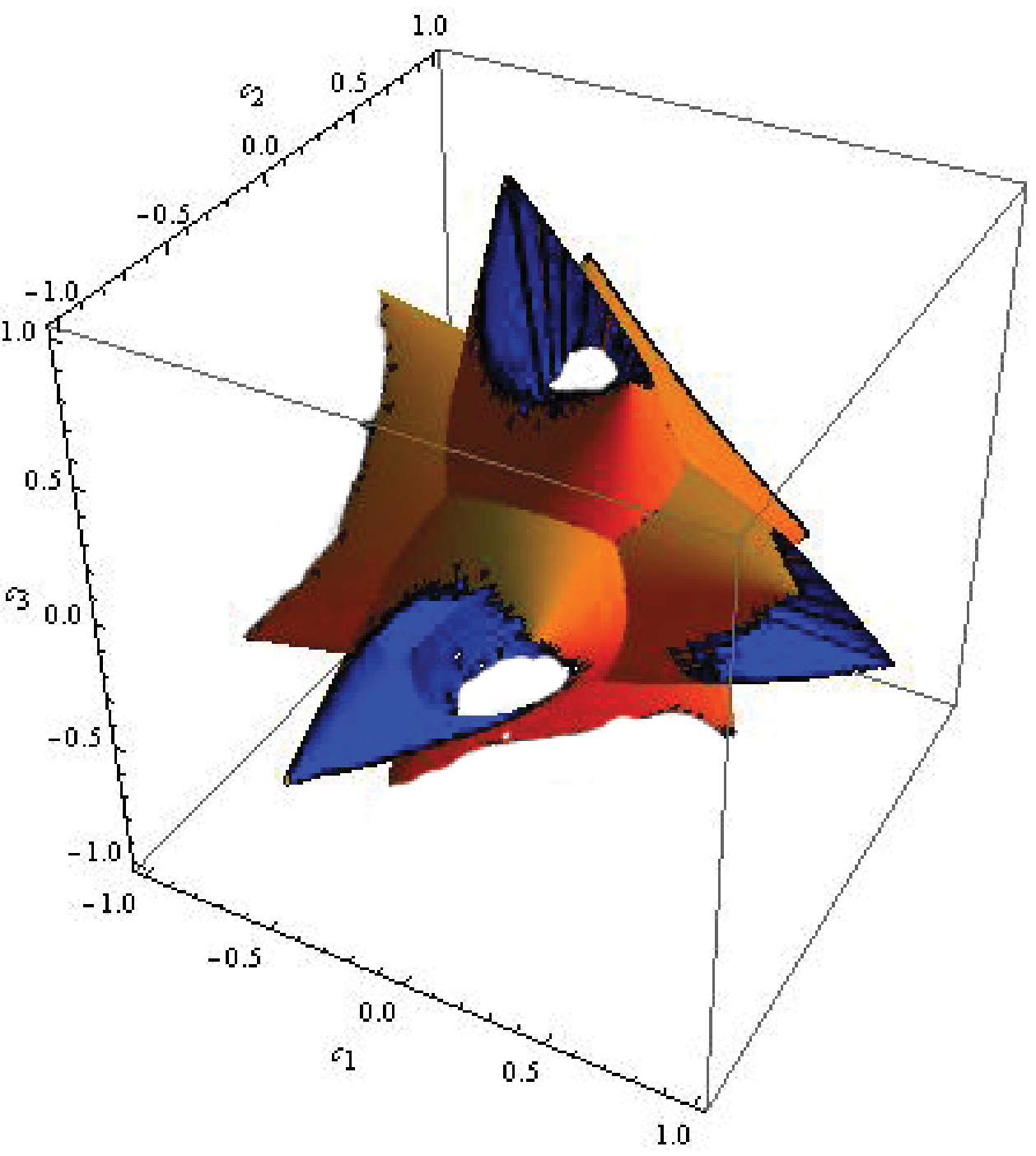}
\qquad
\raisebox{17em}{(d)}\includegraphics[width=6.25cm]{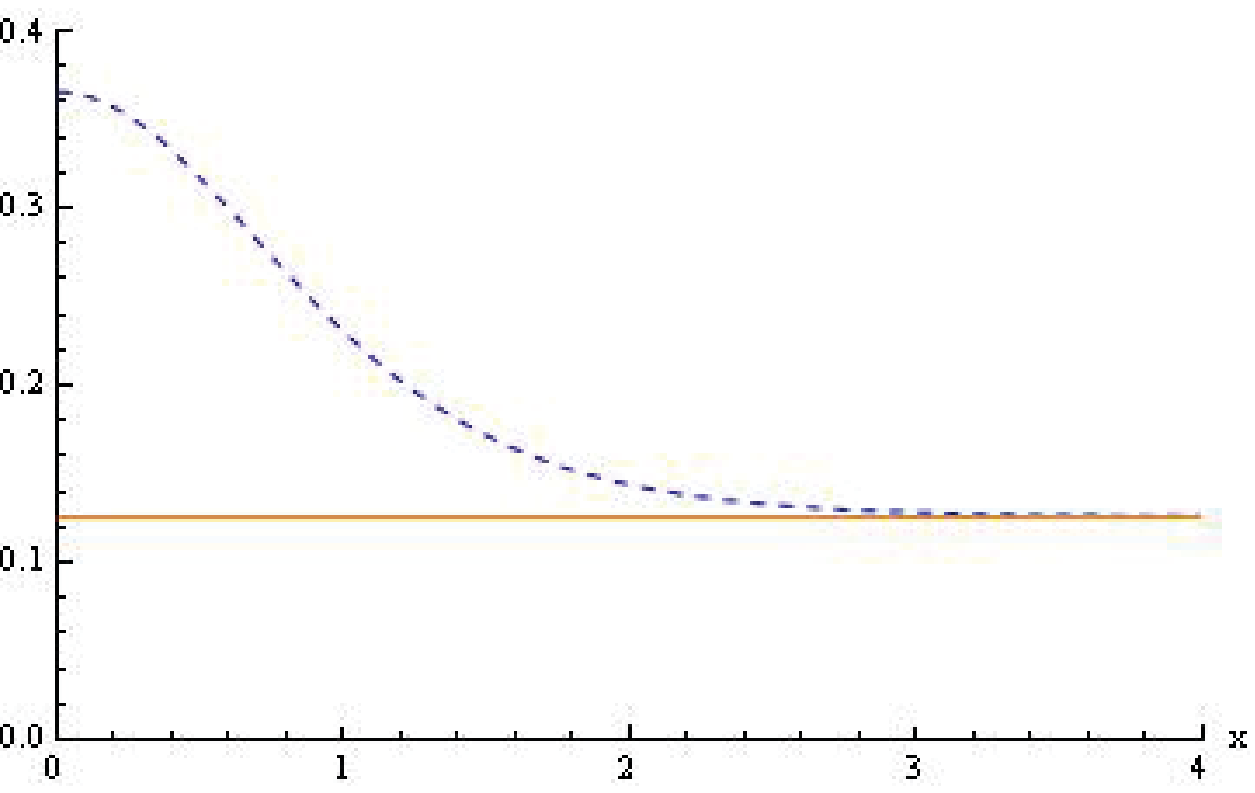}
\caption{(Color online) (a), (b) and (c) are surfaces of constant quantum discord and super-quantum discord for Bell-diagonal state: surface of quantum discord (orange surface), surface of super-quantum discord (blue surface), (a) $D_w=D=0.15$, $x=0.2$; (b) $D_w=D=0.15$, $x=1.5$; (c) $D_w=D=0.15$, $x=3$. (d) super-quantum discord (dashed blue line), quantum discord (solid orange line), $c_{1}=0.3$, $c_{2}=-0.4$, $c_{3}=0.56$.}
\end{center}
\end{figure}

In Fig.2, we plot the surfaces of the super-quantum discord and the quantum discord for the Bell-diagonal state at the same values of parameters in (a),(b) and (c). It is shown that the surface of super-quantum discord is surrounded by the surface of quantum discord for smaller values of $x$. The surfaces of super-quantum discord approaches to the surfaces of quantum discord for larger values of $x$. On the other hand, if we take $c_{1}=0.3$, $c_{2}=-0.4$, $c_{3}=0.56$, then in (d) one finds that the super discord is greater than the normal discord for smaller values of $x$ and approaches to the normal discord for larger values of $x$.

\section{\bf Dynamics of super-quantum correlation and quantum correlation for  the Werner state and Bell-diagonal states under local nondissipative channels}\label{IIII}
In the following, we consider that the Bell-diagonal state undergoes the phase flip channel \cite{Maziero},  with the Kraus operators
$\Gamma_0^{(A)}=$ diag$(\sqrt{1-p/2}, \sqrt{1-p/2})\otimes I$,  $\Gamma_1^{(A)}=$ diag$(\sqrt{p/2}, -\sqrt{p/2})\otimes I$,
$\Gamma_0^{(B)}= I \otimes$ diag$(\sqrt{1-p/2}, \sqrt{1-p/2}) $,  $\Gamma_1^{(B)}= I \otimes$ diag$(\sqrt{p/2}, -\sqrt{p/2}) $,  where $p=1-\exp(-\gamma t)$,  $\gamma$ is
the phase damping rate \cite{Maziero, yu}.

Let $\varepsilon(\cdot)$ represents the operator of decoherence. Then, under the phase flip channel,  we have
\begin{eqnarray}
\varepsilon(\rho_{AB})&=& \frac{1}{4}(I\otimes I+(1-p)^2c_1\sigma_1\otimes\sigma_1\nonumber\\
    &&+(1-p)^2c_2\sigma_2\otimes\sigma_2+c_3\sigma_3\otimes\sigma_3).
\end{eqnarray}
Werner state under the phase flip channel is given by
 \begin{eqnarray}
\varepsilon(\rho_{AB})&=& \frac{1}{4}(I\otimes I-(1-p)^2z\sigma_1\otimes\sigma_1\nonumber\\
    &&-(1-p)^2z\sigma_2\otimes\sigma_2-z\sigma_3\otimes\sigma_3).
\end{eqnarray}

The super-quantum discord of the Werner state under the phase flip channel is given by
 \begin{eqnarray}
ND_w(\rho_{AB})&=&(\frac{1-z+4pz-2p^{2}z}{4})\log(\frac{1-z+4pz-2p^{2}z}{4})\nonumber\\
& &+(\frac{1+3z-4pz+2p^{2}z}{4})\log(\frac{1+3z-4pz+2p^{2}z}{4})\nonumber\\
& &+\frac{1-z}{2}\log(\frac{1-z}{4})+1\nonumber\\
& &-\frac{1+z\tanh x}{2}\log\frac{1+z\tanh x}{2}\nonumber\\
& &-\frac{1-z\tanh x}{2}\log\frac{1-z\tanh x}{2}.
\end{eqnarray}
The quantum discord (one-way deficit) of the Werner state under the phase flip channel is given by
\begin{eqnarray}
ND(\rho_{AB})&=&(\frac{1-z+4pz-2p^{2}z}{4})\log(\frac{1-z+4pz-2p^{2}z}{4})\nonumber\\
& &+(\frac{1+3z-4pz+2p^{2}z}{4})\log(\frac{1+3z-4pz+2p^{2}z}{4})\nonumber\\
& &+\frac{1-z}{2}\log(\frac{1-z}{4})+1-\frac{1+z}{2}\log\frac{1+z}{2}-\frac{1-z}{2}\log\frac{1-z}{2}.
\end{eqnarray}
The weak one-way deficit of the Werner state under the phase flip channel is given by
\begin{eqnarray}
N\Delta^{\rightarrow}_{w}(\rho_{AB})&=&(\frac{1-z+4pz-2p^{2}z}{4})\log(\frac{1-z+4pz-2p^{2}z}{4})\nonumber\\
& &+(\frac{1+3z-4pz+2p^{2}z}{4})\log(\frac{1+3z-4pz+2p^{2}z}{4})\nonumber\\
& &-(\frac{1+z}{4}+\frac{(1-p)^{2}z}{2\cosh x})\log(\frac{1+z}{4}+\frac{(1-p)^{2}z}{2\cosh x})\nonumber\\
& &-(\frac{1+z}{4}-\frac{(1-p)^{2}z}{2\cosh x})\log(\frac{1+z}{4}-\frac{(1-p)^{2}z}{2\cosh x}).
\end{eqnarray}

In Fig.3, as an example, the dynamic behaviors of super-quantum correlation and quantum correlation of the Werner  state under the phase flip channel are depicted for $x=0.5$ and $x=3$. Against the decoherence, we find that super-quantum discord is greater than quantum discord (one-way deficit) and quantum discord (one-way deficit) is greater than weak one-way deficit. Super-quantum correlation including super-quantum discord and weak one-way deficit approaches to quantum correlation including quantum discord (one-way deficit) for larger $x$ under the phase flip channel. As $z$ increases, super-quantum correlation and quantum correlation increase. Then, as $p$ increases, super-quantum correlation and quantum correlation decrease.
\begin{figure}[h]
\raisebox{17em}{(a)}\includegraphics[width=6.25cm]{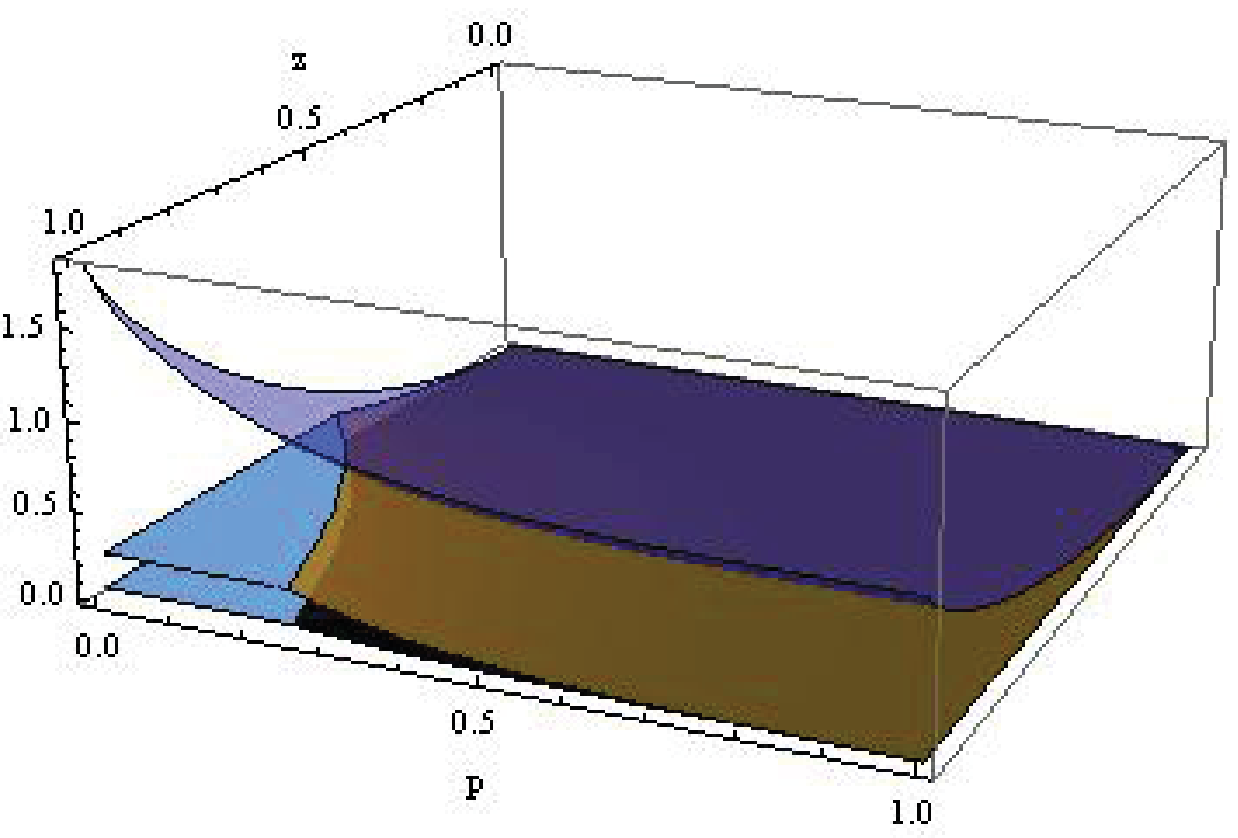}
\raisebox{17em}{(b)}\includegraphics[width=6.25cm]{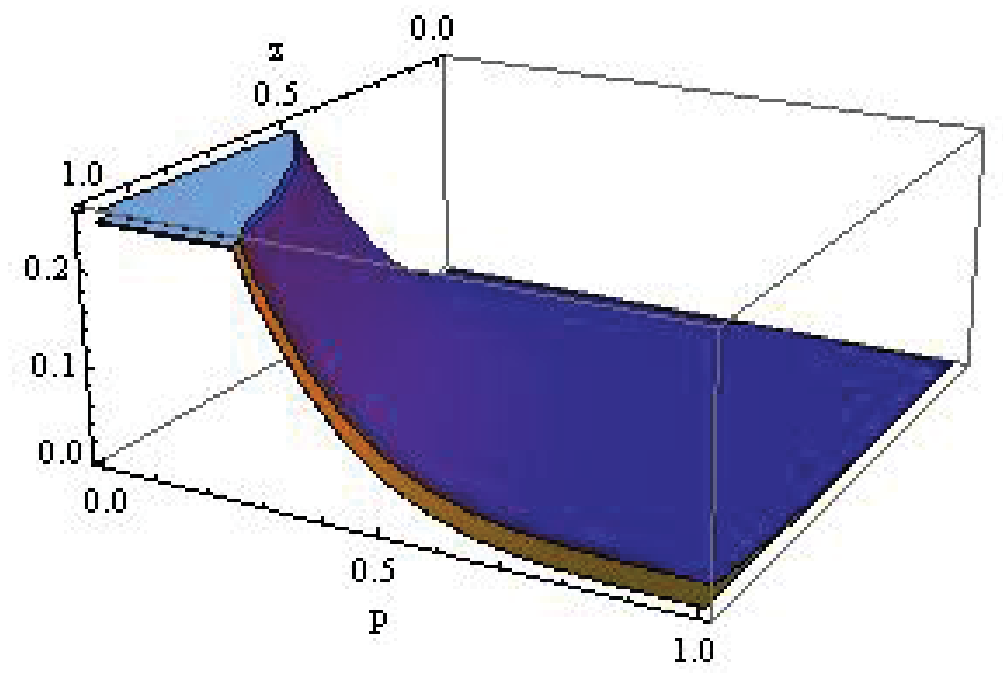}
\label{Fig:3}
\end{figure}
\begin{figure}[h]
\begin{center}
\caption{(Color online) Super-quantum correlation and quantum correlation for the Werner state under the phase flip channel: super-quantum discord (blue surface), quantum discord (one-way deficit) (orange surface) and weak one-way deficit (black surface) as a function of $z$ and $p$: (a) $x=0.5$, (b) $x=3$.}
\end{center}
\end{figure}

 Next, we consider the Bell-diagonal state under the phase flip channel.
Let us assume $|c_{1}|<|c_{2}|<|c_{3}|$. Super-quantum discord and quantum discord for the Bell-diagonal are given by
\begin{eqnarray}
ND_w(\rho_{AB})&=&\frac{1-(1-p)^{2}c_1-(1-p)^{2}c_2-c_3}{4}\log (1-(1-p)^{2}c_1-(1-p)^{2}c_2-c_3)\nonumber\\
& &+\frac{1-(1-p)^{2}c_1+(1-p)^{2}c_2+c_3}{4}\log (1-(1-p)^{2}c_1+(1-p)^{2}c_2+c_3)\nonumber\\
& &+\frac{1+(1-p)^{2}c_1-(1-p)^{2}c_2+c_3}{4}\log (1+(1-p)^{2}c_1-(1-p)^{2}c_2+c_3)\nonumber\\
& &+\frac{1+(1-p)^{2}c_1+(1-p)^{2}c_2-c_3}{4}\log (1+(1-p)^{2}c_1+(1-p)^{2}c_2-c_3)\nonumber\\
& &-\frac{1-c_{3}\tanh x}{2}\log(1-c_{3}\tanh x)-\frac{1+c_{3}\tanh x}{2}\log(1+c_{3}\tanh x),
\end{eqnarray}
\begin{eqnarray}
ND(\rho _{AB})&=&\frac{1-(1-p)^{2}c_1-(1-p)^{2}c_2-c_3}{4}\log (1-(1-p)^{2}c_1-(1-p)^{2}c_2-c_3)\nonumber\\
& &+\frac{1-(1-p)^{2}c_1+(1-p)^{2}c_2+c_3}{4}\log (1-(1-p)^{2}c_1+(1-p)^{2}c_2+c_3)\nonumber\\
& &+\frac{1+(1-p)^{2}c_1-(1-p)^{2}c_2+c_3}{4}\log (1+(1-p)^{2}c_1-(1-p)^{2}c_2+c_3)\nonumber\\
& &+\frac{1+(1-p)^{2}c_1+(1-p)^{2}c_2-c_3}{4}\log (1+(1-p)^{2}c_1+(1-p)^{2}c_2-c_3)\nonumber\\
& &-\frac{1-c_{3}}{2}\log(1-c_{3})-\frac{1+c_{3}}{2}\log(1+c_{3}).
\end{eqnarray}
 We plot in Fig.4 the super-quantum discord and quantum discord for the Bell-diagonal state under the phase flip channel. We find in (a) that super-quantum discord is greater than quantum discord. They approach together for bigger values of $x$.  The super-quantum discord under the phase flip channel as a function of $x$ and $p$ is shown in (b): the super-quantum discord decreases as $x$ and $p$ increase.
\begin{figure}[h]
\raisebox{17em}{(a)}\includegraphics[width=6.25cm]{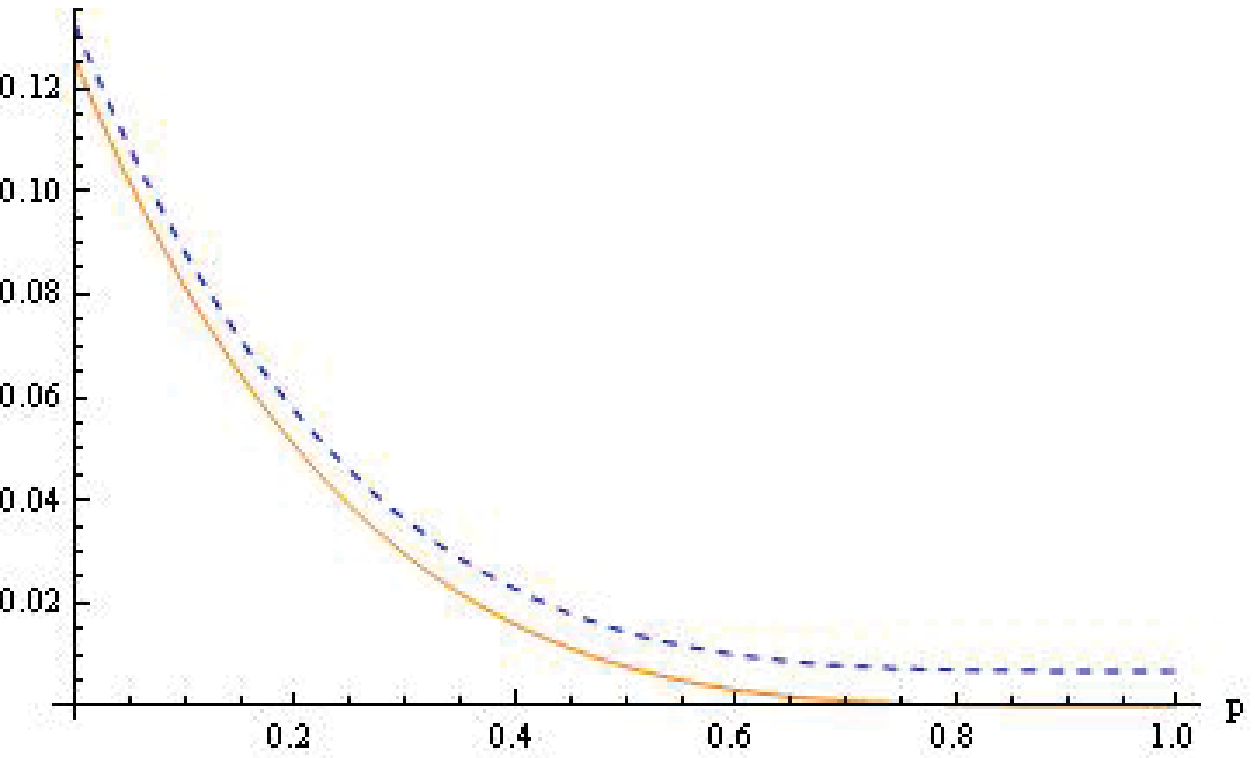}
\raisebox{17em}{(b)}\includegraphics[width=6.25cm]{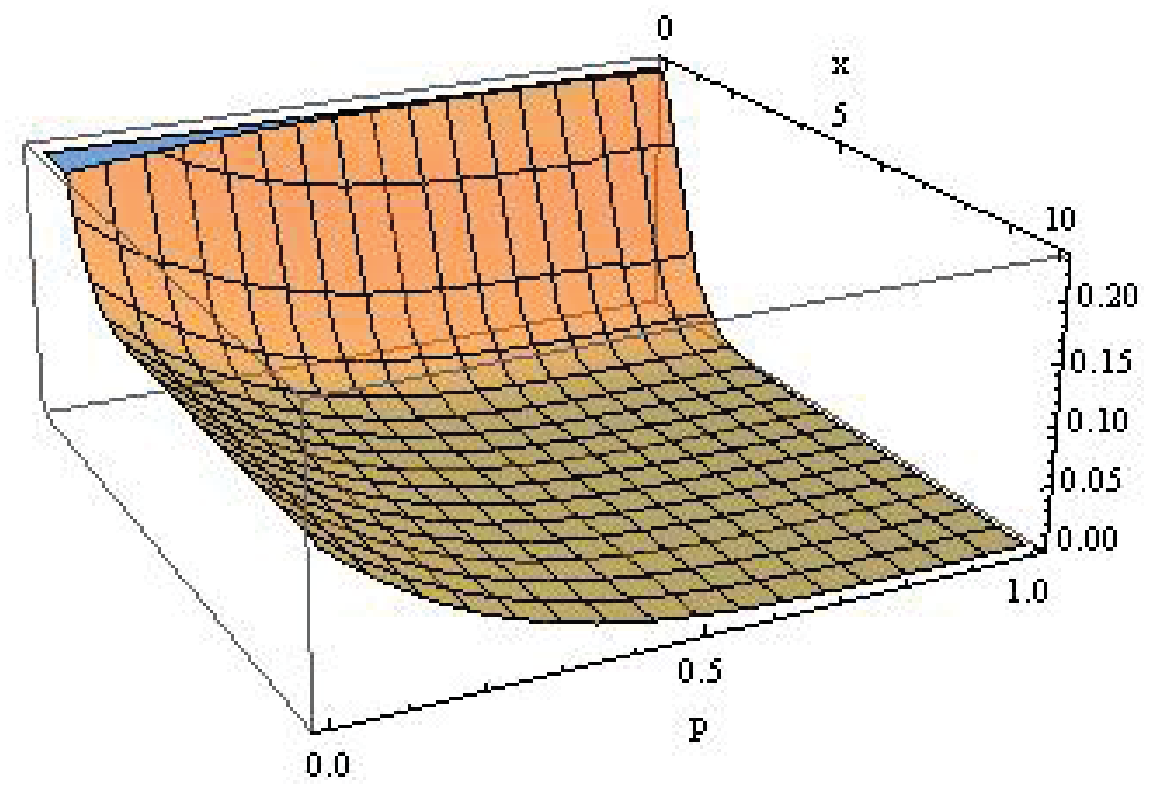}
\label{Fig:4}
\end{figure}
\begin{figure}[h]
\begin{center}
\caption{(Color online) Super-quantum discord and quantum discord for the Bell-Diagonal state under the phase flip channel: (a) super-quantum discord (dashed blue line) and quantum discord (one-way deficit) (solid orange line) as a function of $p$ for $x=2.5,~ c_{1}=0.3,~ c_{2}=-0.4,~ c_{3}=0.56$. (b) Super-quantum discord as a function of $x$ and $p$ for $c_{1}=0.3,~ c_{2}=-0.4,~ c_{3}=0.56$.}
\end{center}
\end{figure}

\section{\bf summary}\label{IIIII}

We have proposed the weak one-way deficit under weak measurement.
The weak one-way deficit has been calculated analytically for Werner state. We find that for Werner states,
while the standard quantum discord and the one-way deficit are the same,
 the weak one-way deficit is smaller than the standard one-way deficit, which contrasts with the fact that
 super-quantum discord is larger than quantum discord. In this sense, for weak measurement, weak one-way deficit
 and super-quantum discord capture different aspects of quantum correlations.
Based on analytic and numerical results, the geometry of super-quantum discord for Bell-diagonal state has been explicitly shown in terms of figures. From the figures,
 we have found that the surface of super-quantum discord is surrounded by the surface of quantum discord for smaller values of $x$ and they approach together for larger values of $x$. The dynamics of quantum correlations for phase flipping channel has been also studied.

The quantification of various correlations is a basic problem for quantum information science. Various measures
provide us different perspectives about quantum and classical correlations. For the weak measurement case, the behaviors of quantum
correlations vary a lot with the strength of the weak measurement. However, it seems
that those measures can still be applied to various protocols in quantum information processing
and identify the importance of the quantum correlations in those protocols.

\bigskip
\noindent {\bf Acknowledgments}  This work was supported by the NSFC11275131 and KZ201210028032.

\end{document}